\documentclass[a4paper,11pt,preprint]{article}
\pdfoutput=1 

\usepackage{jheppub} 

\usepackage[T1]{fontenc} 

\usepackage{graphicx}
\usepackage{amsmath}
\usepackage{amssymb}

\def\ba{\begin{eqnarray}}
\def\ea{\end{eqnarray}}

\def\be{\begin{equation}}
\def\ee{\end{equation}}

\newcommand*\barM{{\overline{\cal M}\hspace{0.5mm}}}

\newcommand{\checked}[1]{}

\title{Exact solution for the non-equilibrium attractor in number-conserving relaxation time approximation}

\author{Michael Strickland} 
\emailAdd{mstrick6@kent.edu}
\author{and Ubaid Tantary} 
\emailAdd{utantary@kent.edu}
\affiliation{Department of Physics, Kent State University, Kent, OH 44242 United States}

\abstract{
We extend previous studies of the conformal 0+1d kinetic non-equilibrium attractor in relaxation time approximation by enforcing number conservation through the introduction of a dynamical fugacity (chemical potential).  We derive two coupled integral equations for the effective temperature and fugacity which are then solved numerically to obtain the exact solution.  The resulting solutions exhibit convergence to a unique non-equilibrium attractor when the scaled moments of the distribution function are plotted as a function of the rescaled time $\overline{w} = \tau/\tau_{\rm eq}$.  This occurs even though the system is out of chemical equilibrium at late times.  In addition, compared to the case where number conservation was not imposed, we find that the moments converge to their respective attractors more quickly, particularly for moments with $m=0$.  Finally, we compare the resulting attractor moments with predictions from different hydrodynamic frameworks.
}

\date{\today}


\begin{document}

\maketitle
\flushbottom

\section{Introduction}
\label{sect:intro}

Based on theory to data comparisons produced over the course of the last decades, there is now a strong body of evidence that the dynamics of the quark-gluon plasma (QGP) created in ultrarelativistic heavy-ion collisions (URHICs) is well described by relativistic dissipative hydrodynamics  \cite{Averbeck:2015jja,Jeon:2016uym,Romatschke:2017ejr,Florkowski:2017olj,Alqahtani:2017mhy}.  Since the phenomenologically extracted values of the shear viscosity to entropy density ratio are finite, this implies that at early times after the nuclear pass through ($0.01-0.1 \lesssim \tau \lesssim 1$ fm/c), the QGP possesses large non-equilibrium corrections, e.g. large pressure anisotropy in the local rest frame.  Because the system experiences large deviations from local thermal equilibrium, one might expect dissipative hydrodynamics approaches to fail at early times.  In practice, however, one finds that dissipative hydrodynamics describes the evolution of the components of the energy-momentum tensor quite well after a rather short time scale $\tau_{\rm hydro} \sim 0.5-1$ fm/c in the center of the overlap region for a central collision.  Since dissipative hydrodynamics frameworks perform well after $\tau \sim \tau_{\rm hydro}$, the system is said to {\em hydrodynamize} at this time scale~\cite{Chesler:2008hg}.  The time scale for hydrodynamization has been extracted by comparing numerical solutions of the underlying microscopic dynamical equations to dissipative hydrodynamics evolution in both the weak and strong coupling limits \cite{Chesler:2008hg,Heller:2013oxa,Keegan:2015avk,Romatschke:2017vte,Strickland:2017kux}.  From these studies one finds that $\tau_{\rm hydro} \sim 2/T$.  At the highest LHC energies and assuming $\eta/s = 0.2$, this translates into $\tau_{\rm hydro} \sim 0.5$ fm/c when considering the center of the fireball created in a zero impact parameter collision.

The fact that the system is quickly driven towards dissipative hydrodynamical evolution can be understood using the concept of the {\em hydrodynamical attractor} \cite{Heller:2015dha}.  In 0+1d conformal viscous hydrodynamics, one can reduce the two coupled equations for the energy density and the shear pressure correction to a single ordinary differential equation which, subject to the correct boundary conditions, provides a universal ``attractor'' solution for the scaled shear correction $\bar\pi = \pi^\eta_\eta/\varepsilon$ as a function of the scaled time $\bar{w} = \tau/\tau_{\rm eq}$, for example.  If one solves the hydrodynamic equations with different initial conditions and plots the results versus $\bar{w}$, one finds that the solutions with different initial conditions converge to the universal attractor solution on a very short time scale (in the sense of small $\bar{w}$).  This observation is not restricted to second-order viscous hydrodynamics and has been shown to hold in numerical solutions to Einstein's equations obtained in the strong coupling limit of ${\cal N}=4$ supersymmetric Yang-Mills in the large $N$ limit \cite{Heller:2013oxa,Keegan:2015avk,Romatschke:2017vte}, QCD effective kinetic theory simulations \cite{Keegan:2015avk,Kurkela:2015qoa,Romatschke:2017vte}, third-order viscous hydrodynamics \cite{JaiswalForth}, anisotropic hydrodynamics \cite{Strickland:2017kux}, and exact solutions to the Boltzmann equation in relaxation time approximation (RTA) subject to both Bjorken and Gubser flows~\cite{Romatschke:2017vte,Strickland:2017kux,Behtash:2017wqg,Behtash:2018moe,Denicol:2018pak,Strickland:2018ayk,Behtash:2019txb}.  

Recently, using the exact solution of the RTA Boltzmann equation subject to Bjorken flow, it was demonstrated that the idea of the non-equilibrium attractor can be extended beyond the low-order moments of the one-particle distribution function typically considered in hydrodynamic approaches~\cite{Strickland:2018ayk}.  In Ref.~\cite{Strickland:2018ayk} it was demonstrated that the full one-particle distribution exhibits attractor-like behavior and that higher moments, ${\cal M}^{nm}$ of the one-particle distribution function converge more quickly to their respective attractors, with the exception being moments with $m=0$, which are more sensitive to the squeezed free-streaming part of the exact solution.  For moments with large $m$ and $n$, Ref.~\cite{Strickland:2018ayk} showed that there is a parametrically large separation between the scaled time at which solutions converge to the non-equilibrium attractor $\bar{w}_c$ and the time at which the moment approaches to within 10\% of its equilibrium value $\bar{w}_{\rm therm}$.  Finally, In Ref.~\cite{Strickland:2018ayk} comparisons were made between the exact attractor moments and various dissipative hydrodynamics frameworks including relativistic Navier-Stokes (NS) \cite{Eckart:1940te,ldlandau2013,Weinberg:1971mx}, second order viscous hydrodynamics~\cite{Muller:1967zza,Israel:1976tn,Israel:1979wp,Muronga:2001zk,Muronga:2003ta,Muronga:2004sf,Heinz:2005bw,Baier:2006um,Romatschke:2007mq,Baier:2007ix,Dusling:2007gi,Luzum:2008cw,Song:2008hj,Heinz:2009xj,Schenke:2010rr,Schenke:2011tv,Bozek:2011wa,Niemi:2011ix,Denicol:2011fa,Niemi:2012ry,Bozek:2012qs,Denicol:2012cn,Denicol:2012es,Denicol:2014vaa,Denicol:2014mca,Jaiswal:2014isa}, third-order viscous hydrodynamics \cite{Jaiswal:2013npa,Jaiswal:2013vta}, and anisotropic hydrodynamics \cite{Florkowski:2010cf,Martinez:2010sc,Ryblewski:2010ch,Florkowski:2011jg,Martinez:2012tu,Ryblewski:2012rr,Bazow:2013ifa,Tinti:2013vba,Nopoush:2014pfa,Florkowski:2014bba,Tinti:2015xwa,Bazow:2015cha,Bazow:2015zca,Nopoush:2015yga,Alqahtani:2015qja,Molnar:2016vvu,Molnar:2016gwq,Bluhm:2015raa,Bluhm:2015bzi,Alqahtani:2017jwl,Alqahtani:2017tnq,Alqahtani:2017mhy,Almaalol:2018ynx,Almaalol:2018gjh}.  It was found that in all cases anisotropic hydrodynamics provided the best approximation to the exact attractor irregardless of the moment considered.  

Importantly, it was shown that, when $m$ or $n$ are large, both the Navier-Stokes and second order viscous hydrodynamics results for the attractor failed to describe the exact solution.  The fact that a subset of the exact moment solutions converge to something that is not well-described by traditional viscous hydrodynamics treatments means that we must refine our terminology a bit:  instead of calling the convergence to the attractor ``hydrodynamization'', we should instead call it {\em pseudo-thermalization} to emphasize that the attractor has a non-hydrodynamic nature reflected in the behavior of higher moments of the one-particle distribution function.  In addition, we can associate the loss of memory of the precise initial conditions used with the pseudo-thermalization of the system.  This is similar to the loss of memory which occurs if a system fully thermalizes, but with the universal state which emerges after pseudo-thermalization being far from equilibrium.

In this paper, we extend Ref.~\cite{Strickland:2018ayk} to study the effect of imposing number conservation on the dynamics and underlying non-equilibrium attractor.  In RTA, one can enforce number conservation by introducing a fugacity (chemical potential) in both the dynamical and equilibrium distribution functions \cite{Florkowski:2012as,Florkowski:2015cba,Florkowski:2017ovw,Almaalol:2018jmz}.  Requiring both energy and number conservation, one can derive two coupled integral equations which can be solved iteratively in order to obtain the effective temperature $T$ and fugacity $\Gamma$ as a function of proper-time.  We demonstrate that for classical statistics, the integral equations can be written solely in terms of the rescaled variables $\hat{T} = T/T_0$ and $\hat{\Gamma} = \Gamma/\Gamma_0$ and the initial momentum space anisotropy $\xi_0$.  As a result, one can construct the exact solution from any initial temperature and fugacity from a trivial scaling of the solution obtained for $\hat{T}$ and $\hat{\Gamma}$.  We then determine the attractor solution to the coupled integral equations numerically by finding solutions which obey $\lim_{\tau \rightarrow 0} {\cal P}_L/{\cal P}_T \rightarrow 0$.  We find that, in general, the resulting attractor solutions do not reach chemical equilibrium at late times, i.e. $\lim_{\tau \rightarrow \infty} \Gamma(\tau) \neq 1$.  Despite the existence of a finite chemical potential at late times, we still observe attractor behavior in all moments and the full distribution function itself.

The structure of this paper is as follows.  In Sec.~\ref{sect:setup}, we briefly review how to rewrite the 0+1d RTA Boltzmann equation using boost invariant variables.  In Sec.~\ref{sect:sol} we present the integral equations obeyed by the one-particle distribution function and all moments of the one-particle distribution function.  In Sec.~\ref{sect:results}, we present our numerical results and discussion of the results.  In Sec.~\ref{sect:conclusions}, we present our conclusions and an outlook for the future.

\section{Setup}
\label{sect:setup}

Our starting point is the Boltzmann equation for massless particles
\be
 p^\mu \partial_\mu  f(x,p) =  C[ f(x,p)] \, , 
\label{kineq}
\ee
\checked{um}
in RTA,
\ba
C[f] = - \frac{p \cdot u}{\tau_{\rm eq}} \left( f-f_{\rm eq} \right) .
\label{col-term}
\ea
\checked{um}
The relaxation time $\tau_{\rm eq}$ above can depend on proper time, however, since the system is conformal (massless) it must be proportional to the inverse effective temperature.   The equilibrium distribution function $f_{\rm eq}$ may be taken to be a Bose-Einstein, Fermi-Dirac, or Boltzmann distribution.  Here we will assume that $f_{\rm eq}$ is given by a Boltzmann distribution
\ba
f_{\rm eq}(\tau,p) = \Gamma(\tau) \exp\left(- \frac{p \cdot u(\tau)}{T(\tau)} \right).
\label{Boltzmann}
\ea
\checked{um}
where $\Gamma(\tau) = \exp(-\mu_{\rm eff}(\tau)/T(\tau))$ is the effective fugacity with $\mu_{\rm eff}(\tau)$ and $T(\tau)$ being the local effective chemical potential and temperature, respectively.

The effective temperature $T$ and fugacity $\Gamma$ will be obtained via matching conditions which demand that the energy and number densities calculated from the dynamical distribution function $f$ be equal to the energy and number densities determined from the equilibrium distribution, $f_{\rm eq}$.  The quantity $u^\mu$ represents the four-velocity of the local rest frame of the matter which herein we assume to be given by the transversally homogenous and boost invariant Bjoken flow (0+1d).

In equilibrium, for massless particles obeying classical statistics the particle density, entropy density, energy density, and pressure are
\ba
n_{\rm eq} = \frac{\Gamma T^3}{\pi^2} \, , \quad
{\cal S}_{\rm eq} = \frac{4 \Gamma T^3}{\pi^2} \, ,
\nonumber \\
{\cal E}_{\rm eq} = \frac{3 \Gamma T^4}{\pi^2} \, , \quad
{\cal P}_{\rm eq} = \frac{\Gamma T^4}{\pi^2} \, .
\label{eq-therm}
\ea
\checked{um}

\subsection{Boost-invariant variables}
\label{sect:boostinvvar}

For one-dimensional boost-invariant expansion, all scalar functions of time and space depend only on the longitudinal proper time $\tau \equiv \sqrt{t^2-z^2}$.  In addition, the hydrodynamic flow $u^\mu$ has the following form $u^\mu = \left(\frac{t}{\tau},0,0,\frac{z}{\tau}\right)$~\cite{Bjorken:1982qr}.  One can introduce a space-like vector that is orthogonal in all frames and corresponds to the z-direction in the local rest frame of the matter \mbox{$z^\mu = \left(\frac{z}{\tau},0,0,\frac{t}{\tau}\right)$}.  The requirement of boost invariance implies that $f(x,p)$ can depend only on three variables: $\tau$, $w$ and $\vec{p}_T$ \cite{Bialas:1984wv,Bialas:1987en,Florkowski:2013lza,Florkowski:2013lya}.  The boost-invariant variable $w$ is defined by
\be
w \equiv  t p_z - z E \, ,
\label{w}
\ee
\checked{um}
where $z$ is the spatial coordinate, not to be confused with the basis vector $z^\mu$. Using $w$ and $\vec{p}_T$ one can define 
\be
v \equiv Et-p_z z = 
\sqrt{w^2+\left( m^2+\vec{p}_T^{\,\,2}\right) \tau^2} \, .  
\label{v}
\ee
\checked{um}
From (\ref{w}) and (\ref{v}) one can easily find the energy and the longitudinal momentum of a particle 
\be
E= p^0 = \frac{vt+wz}{\tau^2} \, ,\quad p_z=\frac{wt+vz}{\tau^2} \, .  
\label{p0p3}
\ee
\checked{um}
The momentum integration measure in phase-space is 
\be
dP =   \frac{d^4p}{(2\pi)^4} \, 2\pi \delta \left( p^2-m^2\right) 2 \theta (p^0)
=\frac{dp_z}{(2\pi)^3p^0}d^2p_T =\frac{dw \, d^2p_T }{(2\pi)^3v}\, .  
\label{dP}
\ee
\checked{um}
In the following we shall consider massless partons, $m=0$.

\subsection{Boost-invariant form of the kinetic equation}
\label{sect:binvkineq}

Making use of the boost-invariant variables introduced in the previous subsection, one finds $p^\mu \partial_\mu f = \frac{v}{\tau} \frac{\partial f}{\partial \tau}$, $p \cdot u = \frac{v}{\tau}$, and $p \cdot z = - \frac{w}{\tau}$.  With this, Eq.~(\ref{kineq}) becomes simply~\cite{Florkowski:2013lza,Florkowski:2013lya}
\ba
\frac{\partial f(\tau,w,p_T)}{\partial \tau}  &=& 
\frac{f_{\rm eq}(\tau,w,p_T)-f(\tau,w,p_T)}{\tau_{\rm eq}(\tau)} \, ,
\label{eq:boltzwvvar}
\ea
\checked{um} 
with the finite chemical potential equilibrium distribution function \eqref{Boltzmann} given by
\ba
f_{\rm eq}(\tau,w,p_T) =\Gamma(\tau)
\exp\!\left[
- \frac{\sqrt{w^2+p_T^2 \tau^2}}{T(\tau) \tau}  \right] .
\label{eqdistform}
\ea
\checked{um}
Note also that $f(\tau,w,\vec{p}_T)$  is an even function of $w$ and depends only on the magnitude of the transverse momentum $\vec{p}_T$.

\section{Exact solution for the distribution function}
\label{sect:sol}

The formal solution of the kinetic equation (\ref{eq:boltzwvvar}) is~\cite{Florkowski:2013lza,Florkowski:2013lya}
\be
f(\tau,w,p_T) = D(\tau,\tau_0)  f_0(w,p_T)  \label{solG} 
+  \int_{\tau_0}^\tau \frac{d\tau^\prime}{\tau_{\rm eq}(\tau^\prime)} \, D(\tau,\tau^\prime) \, 
f_{\rm eq}(\tau^\prime,w,p_T) \, ,  
\ee
\checked{um}
where we have introduced the damping function
\ba
D(\tau_2,\tau_1) = \exp\left[-\int\limits_{\tau_1}^{\tau_2}
\frac{d\tau^{\prime\prime}}{\tau_{\rm eq}(\tau^{\prime\prime})} \right] .
\ea
\checked{um}
For \mbox{$\tau=\tau_0$} the distribution function $f$ reduces to the initial distribution function, $f_0$.   For the conformal RTA solution we use the relation
\ba
\tau_{\rm eq}(\tau) = \frac{5 {\bar \eta}}{T(\tau)} \, ,
\label{taueq}
\ea
\checked{um}
where ${\bar \eta} \equiv \eta/{\cal S}$ is the ratio of the shear viscosity $\eta$ to entropy density ${\cal S}$.

\subsection{Initial distribution}
\label{sect:initdistr}

Herein, for the initial condition we take the Romatschke-Strickland form \cite{Romatschke:2003ms} with a classical Boltzmann distribution as the underlying isotropic distribution 
\be
f_0(w,p_T) =  \gamma_0
\exp\left[
-\frac{\sqrt{(1+\xi_0) w^2 + p_T^2 \tau_0^2}}{\Lambda_0 \tau_0}\, \right] = \gamma_0
\exp\left[
-\frac{\sqrt{ w^2 \alpha_0^{-2}  + p_T^2 \tau_0^2}}{\Lambda_0 \tau_0}\, \right] ,
\label{RS}
\ee
\checked{um}
where, in the second equality, we have introduced the elliptical anisotropy parameter \mbox{$\alpha \equiv 1/\sqrt{1+\xi}$} for convenience.  The distribution function above reduces to an isotropic Boltzmann distribution if the anisotropy parameter $\alpha_0=\alpha(\tau_0)=1$. If $\alpha_0 = 1$, the transverse momentum scale $\Lambda_0$ is equal to the system's initial temperature $T_0$ and the initial microscopic fugacity $\gamma_0$ is equal to the initial effective fugacity $\Gamma_0$.  In general, one must use Landau matching of the initial energy and number densities to fix $\Lambda_0$ and $\gamma_0$ in terms of $T_0$ and $\Gamma_0$.  The resulting ``matching conditions'' are~\cite{Almaalol:2018jmz}
\ba
T &=& \frac{{\cal H}(\alpha)}{2\alpha} \Lambda \, , \label{tmatch} \\
\Gamma &=& \frac{8 \gamma \alpha^4}{{\cal H}^3(\alpha)} \, , \label{gmatch}
\ea
\checked{um}
where ${\cal H}(\alpha) \equiv {\cal H}^{20}(\alpha)$.  The special functions ${\cal H}^{nm}$ needed are
\be
{\cal H}^{nm}(y) \equiv \tfrac{2y^{2m+1}}{2m+1}  {}_2F_1(\tfrac{1}{2}+m,\tfrac{1-n}{2};\tfrac{3}{2}+m;1-y^2)  \, .
\ee
\checked{um}

\subsection{General moments of the distribution function}
\label{sect:genmomevol}

In order to solve Eq.~\eqref{solG}, one can take a general moment of both sides using
\be
{\cal M}^{nm}[f] \equiv \int dP \,(p \cdot u)^n \, (p \cdot z)^{2m} \, f(\tau,w,p_T) \, .
\ee
\checked{um}
For $n=2$ and $m=0$, one obtains the energy density
\be
{\cal E} = {\cal M}^{20} = \int dP \, (p \cdot u)^2 \, f(\tau,w,p_T)  = T^{00}_{\rm LRF} \, ,
\ee
\checked{um}
and, for $n=1$ and $m=0$, one obtains the number density
\be
n = {\cal M}^{10} = \int dP \, (p \cdot u) \, f(\tau,w,p_T) \, ,
\ee
\checked{um}
Using the mass shell constraint, one can always rewrite the transverse momentum squared in terms of the energy and longitudinal momentum, so that any moment containing $p_T^{2\ell}$ can be written as a linear combination of the ${\cal M}^{nm}$ moments above.  As a result, in the general case, we need to compute
\be
{\cal M}^{nm}[f] = \frac{1}{(2\pi)^3 \, \tau^{n+2m}} \int  dw \, d^2p_T  \, v^{n-1} w^{2m} \, f(\tau,w,p_T)
\ee
\checked{um}

\subsection{Integral equation obeyed by a general moment}

Taking a general moment of Eq.~(\ref{solG}) one obtains
\be
{\cal M}^{nm}(\tau) = D(\tau,\tau_0) {\cal M}^{nm}_0(\tau)  + \int_{\tau_0}^\tau \frac{d\tau^\prime}{\tau_{\rm eq}(\tau^\prime)} \, D(\tau,\tau^\prime) \, 
{\cal M}^{nm}_{\rm eq}(\tau,\tau') \, ,
\ee
\checked{um}
where
\ba
{\cal M}^{nm}_0(\tau) &=& \frac{\gamma_0  \, (n+2m+1)! \, \Lambda_0^{n+2m+2}}{(2\pi)^2} {\cal H}^{nm}\left( \frac{\alpha_0 \tau_0}{\tau} \right) , \\
{\cal M}^{nm}_{\rm eq}(\tau,\tau') &=& \frac{\Gamma(\tau')  \, (n+2m+1)! \, T^{n+2m+2}(\tau')}{(2\pi)^2} {\cal H}^{nm} \hspace{-1mm} \left( \frac{\tau'}{\tau} \right) .
\ea
\checked{um}

One can rewrite the first term, which involves the initial values of the microscopic parameters $\gamma_0$ and $\Lambda_0$, in terms of the initial effective fugacity $\Gamma_0$ and temperature $T_0$ using Eqs.~\eqref{tmatch} and \eqref{gmatch}.  Putting the pieces together, our final result for the general moment equation is
\ba
{\cal M}^{nm}(\tau) &=& \frac{(n+2m+1)!}{(2\pi)^2} \Bigg[ D(\tau,\tau_0) \alpha_0^{n+2m-2}  T_0^{n+2m+2} \Gamma_0 \frac{{\cal H}^{nm}\!\left( \frac{\alpha_0 \tau_0}{\tau} \right)}{[{\cal H}(\alpha_0)/2]^{n+2m-1}} \nonumber \\
&& \hspace{3cm} + \int_{\tau_0}^\tau \frac{d\tau^\prime}{\tau_{\rm eq}(\tau^\prime)} \, D(\tau,\tau^\prime) \, 
\Gamma(\tau^\prime)   T^{n+2m+2}(\tau') {\cal H}^{nm} \hspace{-1mm} \left( \frac{\tau'}{\tau} \right) \Bigg] . \hspace{1cm}
\label{eq:meqfinal}
\ea
\checked{um}

\subsection*{Final equations}

From Eq.~\eqref{eq:meqfinal} we can obtain two integral equations by evaluating the $n=1$ and $m=0$ and $n=2$ and $m=0$ moments which map to the number density and energy density, respectively, with the results being
\be
\Gamma(\tau) T^4(\tau) = D(\tau,\tau_0) \Gamma_0 T_0^4 \frac{{\cal H}\!\left( \frac{\alpha_0 \tau_0}{\tau} \right)}{{\cal H}(\alpha_0)} \\
+ \int_{\tau_0}^\tau \frac{d\tau^\prime}{2 \tau_{\rm eq}(\tau^\prime)} \, D(\tau,\tau^\prime) \, 
\Gamma(\tau^\prime) T^4(\tau') {\cal H}\hspace{-1mm} \left( \frac{\tau'}{\tau} \right)  ,
\label{eq:inteq1}
\ee
\checked{um}
and
\be
\Gamma(\tau) T^3(\tau) =  \frac{1}{\tau} \left[ D(\tau,\tau_0) \Gamma_0 T_0^3 \tau_0 + \int_{\tau_0}^\tau \frac{d\tau^\prime}{ \tau_{\rm eq}(\tau^\prime)} \, D(\tau,\tau^\prime) \, 
\Gamma(\tau^\prime) T^3(\tau') \tau'  \right] ,
\label{eq:inteq2}
\ee
\checked{um}
where we used the matching conditions ${\cal E} = {\cal E}_{\rm eq}(T,\Gamma)$ and $n = n_{\rm eq}(T,\Gamma)$ on the left-hand-side and the fact that ${\cal H}^{10}(\alpha) = 2\alpha$ to simplify the second integral equation.  Note, importantly, that one can divide the left- and right-hand sides of Eqs.~\eqref{eq:inteq1} and \eqref{eq:inteq2} by the initial fugacity $\Gamma_0$ and rewrite them entirely in terms of $\hat\Gamma \equiv \Gamma/\Gamma_0$.  As a result, one can solve these coupled integral equations with a given value of $\Gamma_0$, e.g. $\Gamma_0 = 1$, and then obtain solutions with different initial fugacity by scaling the result by the initial fugacity.\footnote{A similar scaling can be done with $T_0$, however, in addition to scaling Eqs.~\eqref{eq:inteq1} and \eqref{eq:inteq2} by $T_0^4$ and $T_0^3$, respectively, one must also make a change of variables in the proper-time integrations by introducing $\hat\tau \equiv \tau T_0$.}

\section{Results}
\label{sect:results}

For our results, we solve Eqs.~\eqref{eq:inteq1} and \eqref{eq:inteq2} numerically to obtain $T(\tau)$ and $\Gamma(\tau)$ given a set of initial values at $\tau_0$:  $T_0$, $\Gamma_0$, and $\alpha_0$.  For this purpose we wrote a CUDA-based GPU code which allows one to efficiently solve Eqs.~\eqref{eq:inteq1} and \eqref{eq:inteq2} on very large proper-time lattices using a logarithmically-spaced grid.  For all results presented herein we used a temporal lattice size of 4096 points and iterated the coupled integral equations until the effective temperature and fugacity converged to sixteen digits at all proper times.  The code used to produce our results is included in the arXiv bundle for this paper and is also publicly available for download using the link provided in Ref.~\cite{MikeCodeDB}.  Once the solutions for the effective temperature and fugacity are obtained, one can use Eq.~\eqref{eq:meqfinal} to obtain the proper-time dependence of any moment required.  One can also use Eq.~\eqref{solG} to reconstruct the full one-particle distribution in a grid in momentum-space.  

We will present the resulting exact solutions for the {\em scaled moments}
\be
\barM^{nm}(\tau) \equiv \frac{{\cal M}^{nm}(\tau)}{{\cal M}^{nm}_{\rm eq}(\tau)} \, ,
\label{mbardef}
\ee
\checked{um}
where
\be
{\cal M}^{nm}_{\rm eq}(\tau) = {\cal M}^{nm}_{\rm eq}(\tau,\tau) = \frac{(n+2m+1)! \,\Gamma(\tau)\, T^{n+2m+2}(\tau)}{2\pi^2 (2m +1)} \, ,
\label{meqres2}
\ee
\checked{um}
are the moments associated with an equilibrium Boltzmann distribution function.  The scaled moments approach one at late times by construction and the rate at which they approach one provides a quantitative measure of how quickly the system thermalizes.  Note that higher moments are sensitive to higher average momenta where the hydrodynamics assumption of small gradients could fail.

\subsection{Attractor moments}

\begin{figure*}[t!]
\centerline{
\includegraphics[width=1\linewidth]{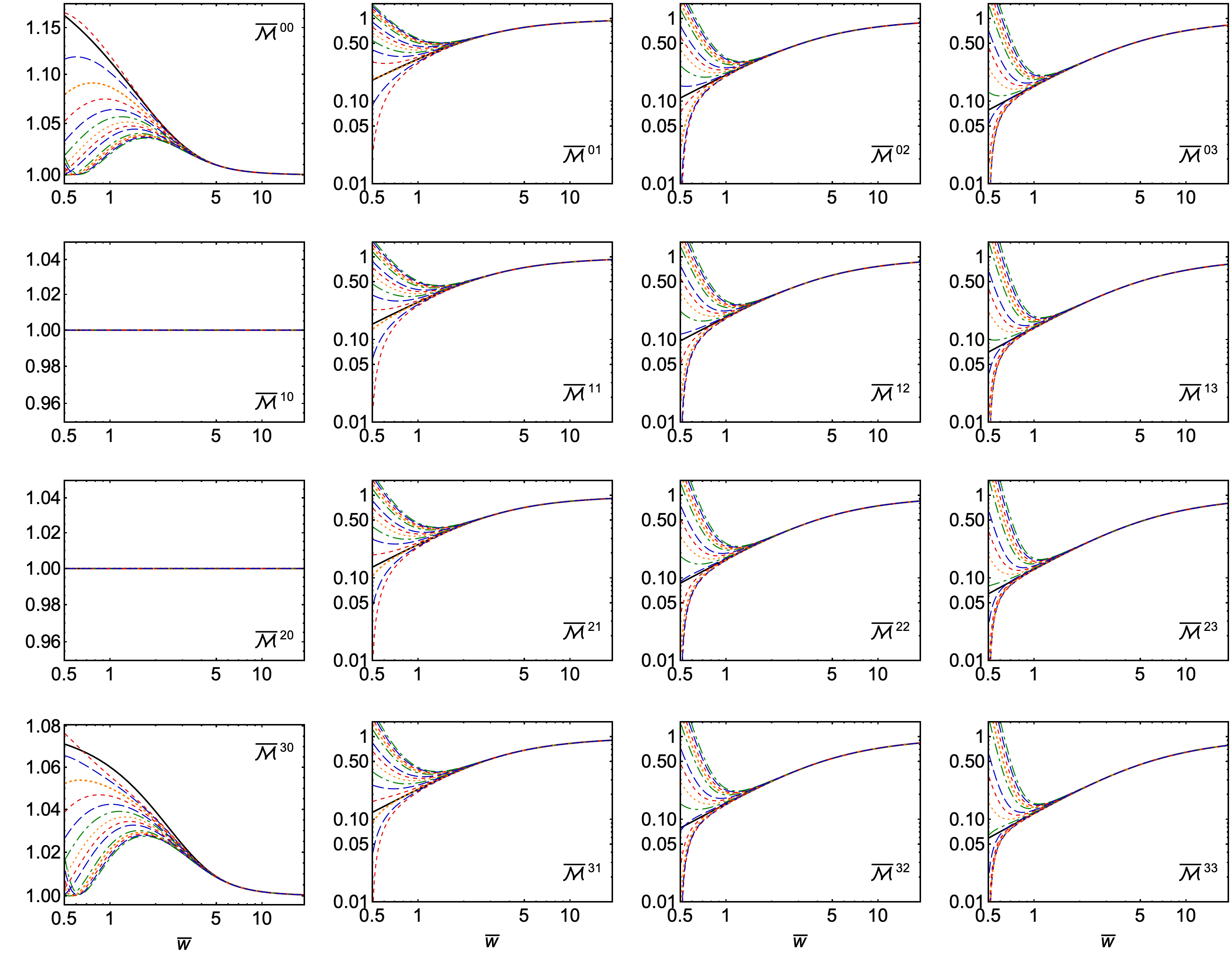}\hspace{2mm}
}
\caption{The scaled moments $\barM^{nm} = {\cal M}^{nm}(\tau)/{\cal M}^{nm}_{\rm eq}(\tau)$ obtained from the exact attractor solution (solid black line) are compared to a set of exact solutions (various colored dotted and dashed lines) initialized with varying $\alpha_0$.  The horizontal axis is $\overline{w} \equiv \tau/\tau_{\rm eq} = \tau T/5 \bar\eta$.  Panels show a grid in $n$ and $m$. }
\label{fig:attractorGridSols}
\end{figure*}
 
In Fig.~\ref{fig:attractorGridSols}, we present sixteen panels containing our numerical results for the scaled moments $\barM^{nm}(\tau)$ with $m,n \in \{ 0,1,2,3 \}$.  In all panels, the horizontal axis  of Fig.~\ref{fig:attractorGridSols} is the scaled proper-time $\overline{w} \equiv \tau/\tau_{\rm eq} = \tau T/5 \bar\eta$, the black line is the exact solution for the attractor, and the various dashed/dotted curves are exact solutions with different values of $\alpha_0$.  To obtain the attractor solution, we solved the coupled integral equations \eqref{eq:inteq1} and \eqref{eq:inteq2} with \mbox{$\tau_0 = 10^{-3}$ fm/c}, $T_0 = 1$ GeV, $\alpha_0 = 2.5 \times 10^{-2}$, and $\Gamma_0 =1$. To obtain the specific solutions (dashed/dotted lines), we solved the same coupled integral equations \eqref{eq:inteq1} and \eqref{eq:inteq2} with $\tau_0 = 10^{-1}$ fm/c, $T_0 = 1$ GeV, $\alpha_0 \in 0.1 \ldots 1.5$, and $\Gamma_0 =1$.  The range of initial anisotropies considered covers both very oblate and very prolate initial momentum-space anisotropy.

As can be seen from this figure, for generic initial conditions, all moments with $m>0$ visually converge to their respective attractors after a short rescaled time $\bar{w} \sim 2$.  For $\barM^{10}$ (scaled number density) and $\barM^{20}$ (scaled energy density), we see that the constraints are properly enforced, resulting in these moments being equal to their equilibrium values at all proper times.  For moments with $m=0$, we see a somewhat slower approach to the attractor.  This is similar to what was found when not enforcing number conservation \cite{Strickland:2018ayk}, however, herein we see smaller deviations from equilibrium.  Despite these smaller deviations from equilibrium compared to the prior studies, moments with $m=0$ still converge more slowly than other moments.  The slow convergence of moments with $m=0$ is related to the fact that they contain no powers of $p_z$ in their integrands and are, therefore, more sensitive to the free streaming term (first term) in Eq.~\eqref{solG}.  Free streaming results in momentum modes from the initial distribution being squeezed to smaller and smaller $|p_z|$ as a function of proper time (see Ref.~\cite{Strickland:2018ayk} for details).

\begin{figure*}[t!]
\centerline{
\includegraphics[width=1\linewidth]{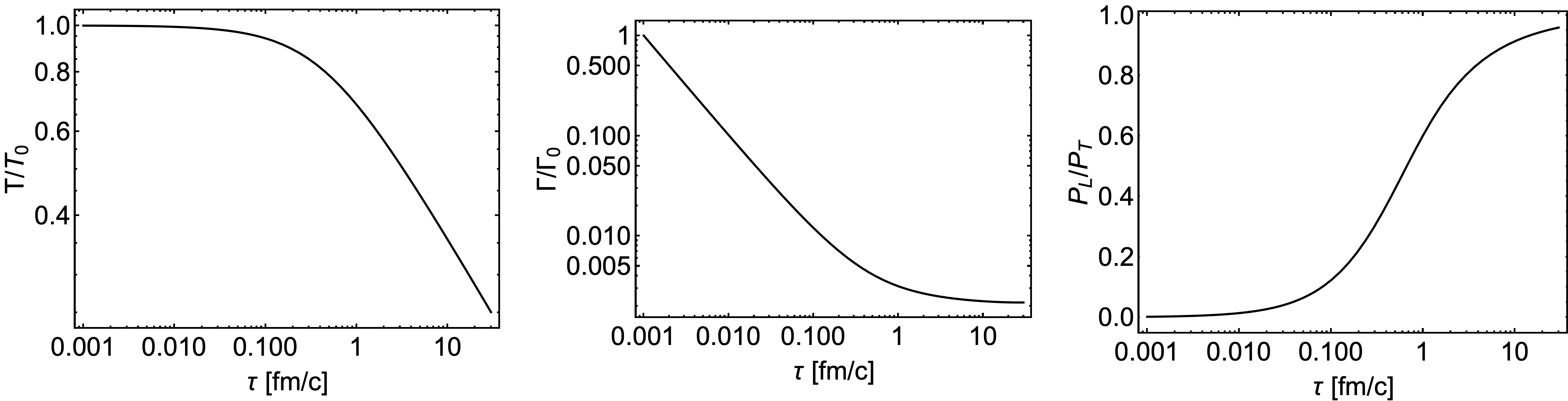}\hspace{2mm}
}
\vspace{-2mm}
\caption{Three panels showing (left) the scaled effective temperature, (middle) the scaled fugacity, and (right) the pressure anisotropy as a function of time for ``attractor'' initial conditions \mbox{$\tau_0 = 10^{-3}$ fm/c}, $T_0 = 1$ GeV, $\alpha_0 = 2.5 \times 10^{-2}$, and $\Gamma_0 =1$.
}
\label{fig:attractorVars}
\end{figure*}

Note, however, that although all scaled moments approach one in the large $\bar{w}$ limit, the system generically possesses a finite fugacity at late times.\footnote{One can adjust the late-time fugacity by changing the initial fugacity $\Gamma_0$.}  To demonstrate this, we plot the effective temperature, effective fugacity, and the pressure anisotropy associated with the attractor in Fig.~\ref{fig:attractorVars}.  As can be seen from this figure, the scaled temperature evolution (left panel) obtained using attractor initial conditions shows characteristics of early time free streaming, for which the temperature scale (average momentum scale) is constant \cite{Mauricio:2007vz,Martinez:2008di,Martinez:2009mf,Martinez:2010sc}, followed by a power law decrease at late time indicative of hydrodynamic evolution.  The attractor's scaled effective fugacity (middle panel) decreases as a power law at early times and eventually saturates at late times.  Finally, we see that attractor's pressure anisotropy is large (highly oblate) at early times with $P_L \ll P_T$ and then slowly relaxes towards isotropy at late times.

\subsection{Pseudo-thermalization time}

In order to quantitatively assess the convergence of generic exact solutions to the attractor for each moment, one can compute the scaled time at which all solutions collapse to the attractor by requiring that $\max|\barM^{nm}_{\rm i}(\overline{w}_{c}) - \barM^{nm}_{\rm attractor}(\overline{w}_{c}) | < \delta_c$ for $i$ in the entire set of trial runs.  In Ref.~\cite{Strickland:2018ayk}, $\delta_c = 10^{-6}$ was chosen in order to require that the solutions were extremely well converged to the attractor.  Herein, we will also consider the weaker convergence criteria of $\delta_c = 10^{-2}$, which should correspond more closely to the time that one extracts when visually checking for convergence in Fig.~\ref{fig:attractorGridSols}.

\begin{figure*}[t!]
\centerline{
\includegraphics[width=0.95\linewidth]{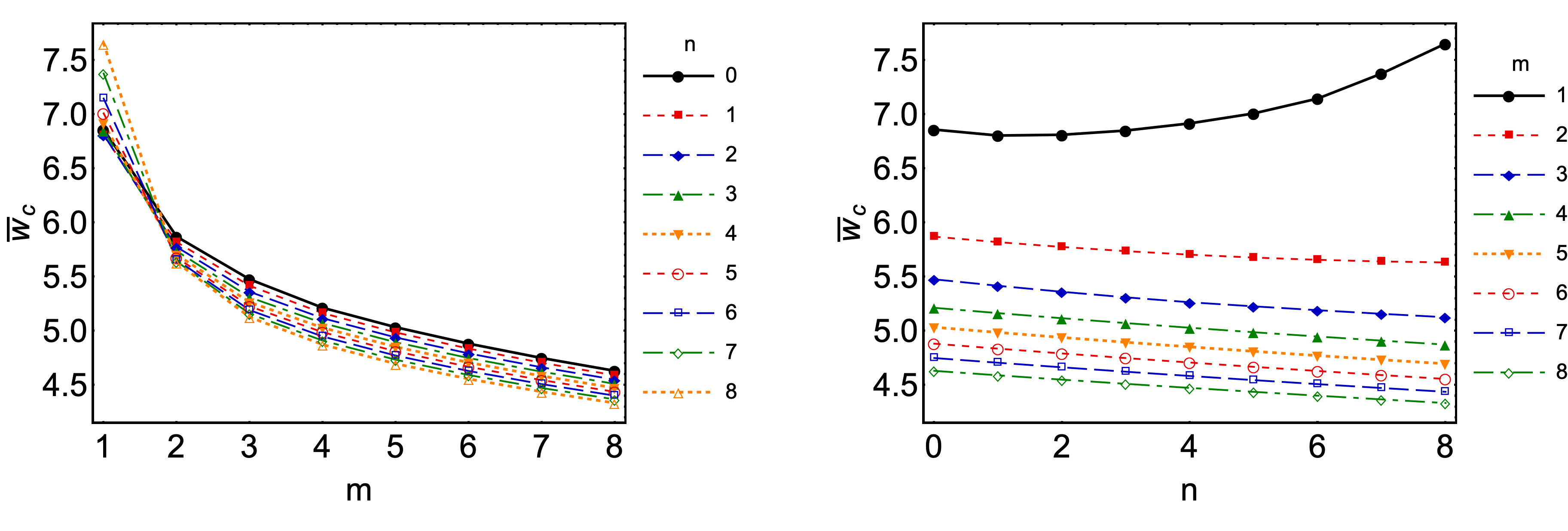}
}
\vspace{4mm}
\centerline{
\includegraphics[width=0.35\linewidth]{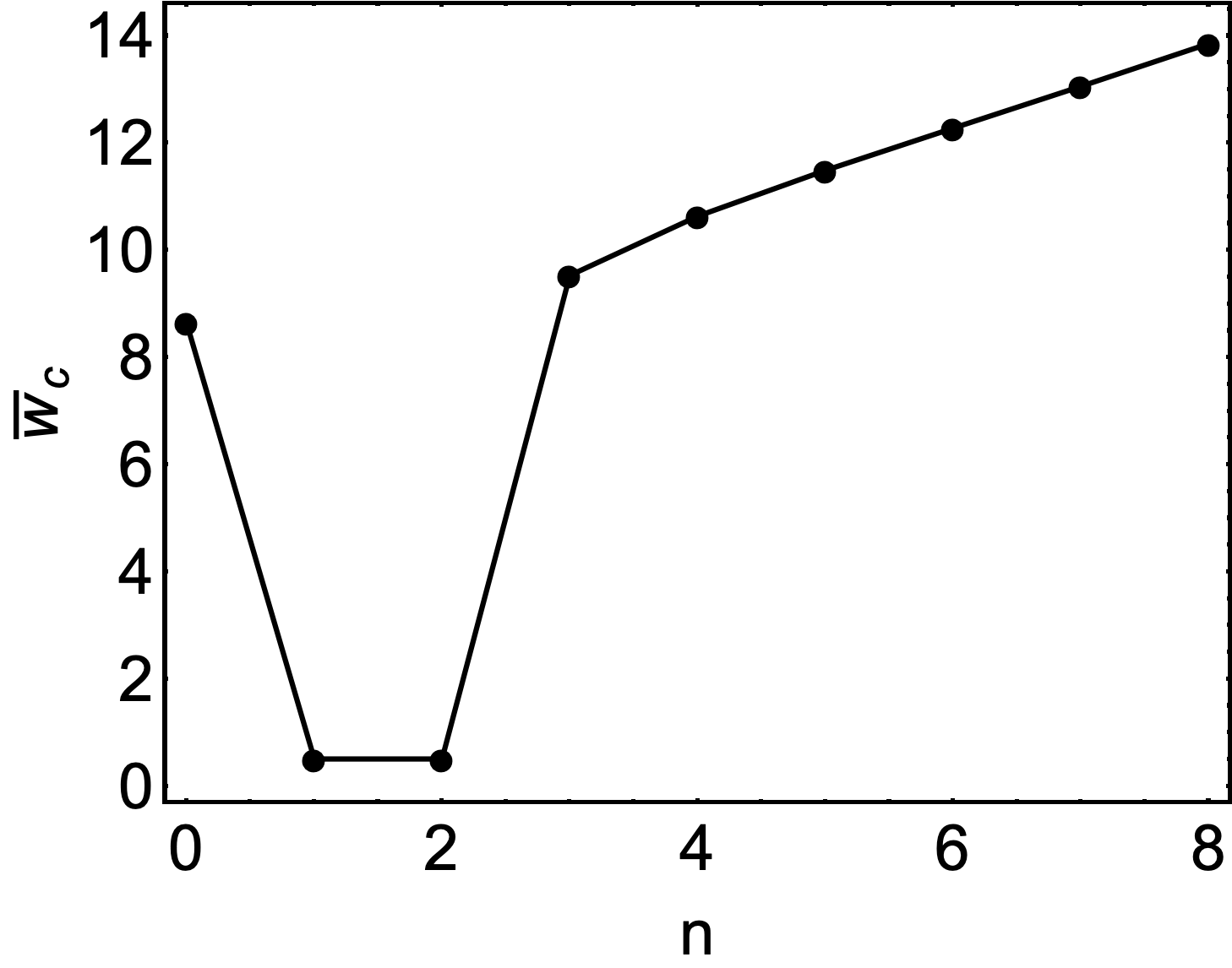}
}
\caption{The pseudo-thermalization time $\overline{w}_c$ for $n,m \in \{0,\cdots,8\}$ and $\delta_c = 10^{-6}$.  The top left panel shows $\overline{w}_c$ as a function of $m$ with the lines corresponding to different values of $n$.  The top right panel shows $\overline{w}_c$ as a function of $n$ with the lines corresponding to different values of $m$.  The bottom panel shows $\overline{w}_c$ as a function of $n$ for the case $m=0$. }
\label{fig:hydrodynamization}
\end{figure*}

\begin{figure*}[t!]
\centerline{
\includegraphics[width=0.95\linewidth]{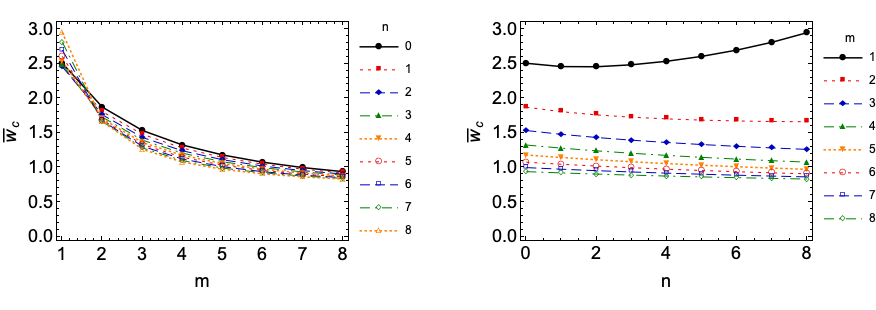}
}
\vspace{4mm}
\centerline{
\includegraphics[width=0.35\linewidth]{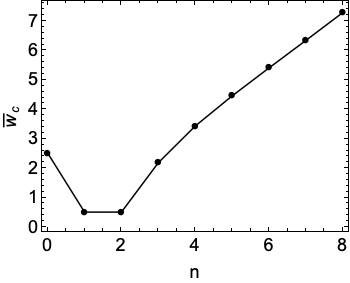}
}
\caption{The pseudo-thermalization time $\overline{w}_c$ for $n,m \in \{0,\cdots,8\}$ and $\delta_c = 10^{-2}$.  Panels are the same as in Fig.~\ref{fig:hydrodynamization}.}
\label{fig:hydrodynamization-2}
\end{figure*}

In Figs.~\ref{fig:hydrodynamization} and \ref{fig:hydrodynamization-2} we plot the convergence or {\em pseudo-thermalization time} $\bar{w}_c$ with $\delta_c = 10^{-6}$ and $\delta_c = 10^{-2}$, respectively.  Figure~\ref{fig:hydrodynamization} uses the strong convergence criteria of $\delta_c = 10^{-6}$ which was the condition used in Ref.~\cite{Strickland:2018ayk}.  The two top panels show $\bar{w}_c$ as a function of $m$ and $n$ and the bottom panel shows the case $m=0$ as a function of $n$.  As can be seen from the top left panel, for $m \geq 2$, $\bar{w}_c$ is a decreasing function of $m$ and $n$.  From the top right panel we see that for $m=1$ the pseudo-thermalization time increases at large $n$, but moments with $m > 1$ have a pseudo-thermalization which decreases as $n$ increases.  Turning to the bottom panel ($m=0$), we see that the $n=1$ and $n=2$ moments thermalize ``instantly'' since these are enforced by conservation laws\footnote{In this case we set $\bar{w}_c$ to be the smallest scaled time in the temporal grid.} and we see a strong increase in $\bar{w}_c$ as $n$ increases.  In the range of $n$ and $m$ shown, the maximum pseudo-thermalization time is $\bar{w}_c^{\rm max} \simeq 14$.  This can be compared with the maximum pseudo-thermalization time obtained without enforcing number conservation (see Fig. 8 in Ref.~\cite{Strickland:2018ayk}), which was $\bar{w}_c^{\rm max} \simeq 28$.  

Turning to Fig.~\ref{fig:hydrodynamization-2} we see the same three panels but now for the weaker convergence criterium of $\delta_c = 0.01$.  As can be seen from these figures, one obtains a shorter pseudo-thermalization time with the weaker convergence criteria, as expected.  In addition, the scaled times extracted for moments with $m \neq 0$ are in the range of $\bar{w}_c \sim 1-3$.  For a RHIC energy heavy-ion collision with a typical initial central temperature of 500 MeV at \mbox{$\tau_0 = 0.1$ fm/c} this translates into a physical pseudo-thermalization time of \mbox{$\tau_c \sim 0.5 - 3$ fm/c} with the precise value depending on the mode considered.  We emphasize that the higher $n$ and $m$ moments converge more quickly and have pseudo-thermalization times on the low side of this window, while the lower $n$ and $m$ moments converge more slowly to the attractor.

\subsection{Comparison with Navier-Stokes, vHydro, and aHydro}

Finally, we compare the exact results for the scaled attractors moments with results obtained from anisotropic hydrodynamics (aHydro) and second-order viscous hydrodynamics (vHydro).  For vHydro, we use the complete second-order viscous hydrodynamics equations of Denicol, Niemi, Molnar, and Rischke (DNMR) \cite{Denicol:2010xn,Denicol:2011fa}.  For aHydro, we use the moments method introduced originally by Florkowski and Tinti \cite{Tinti:2013vba}. For both vHydro and aHydro, the attractor is determined from the solution of a one-dimensional ordinary differential equation subject to the appropriate initial condition.  For details concerning the determination of the attractor for both aHydro and vHydro, we refer to the reader to Ref.~\cite{Strickland:2017kux}.

For vHydro, one extracts $\bar\pi = \pi/\epsilon$ and, using this, one can reconstruct the solution for any moment using~\cite{Strickland:2018ayk}
\be
\barM^{nm}_{\rm vHydro}(\tau) = 1 - \frac{3 m (n+2m+2)(n+2m+3)}{4(2m+3)} \bar\pi \, .
\label{eq:vhydromoms}
\ee
\checked{um}
One can obtain the Navier-Stokes (NS) result by taking $\bar\pi = 16 \bar\eta/(9 \tau T)$ in \eqref{eq:vhydromoms}.  

\begin{figure*}[t!]
\centerline{
\includegraphics[width=1\linewidth]{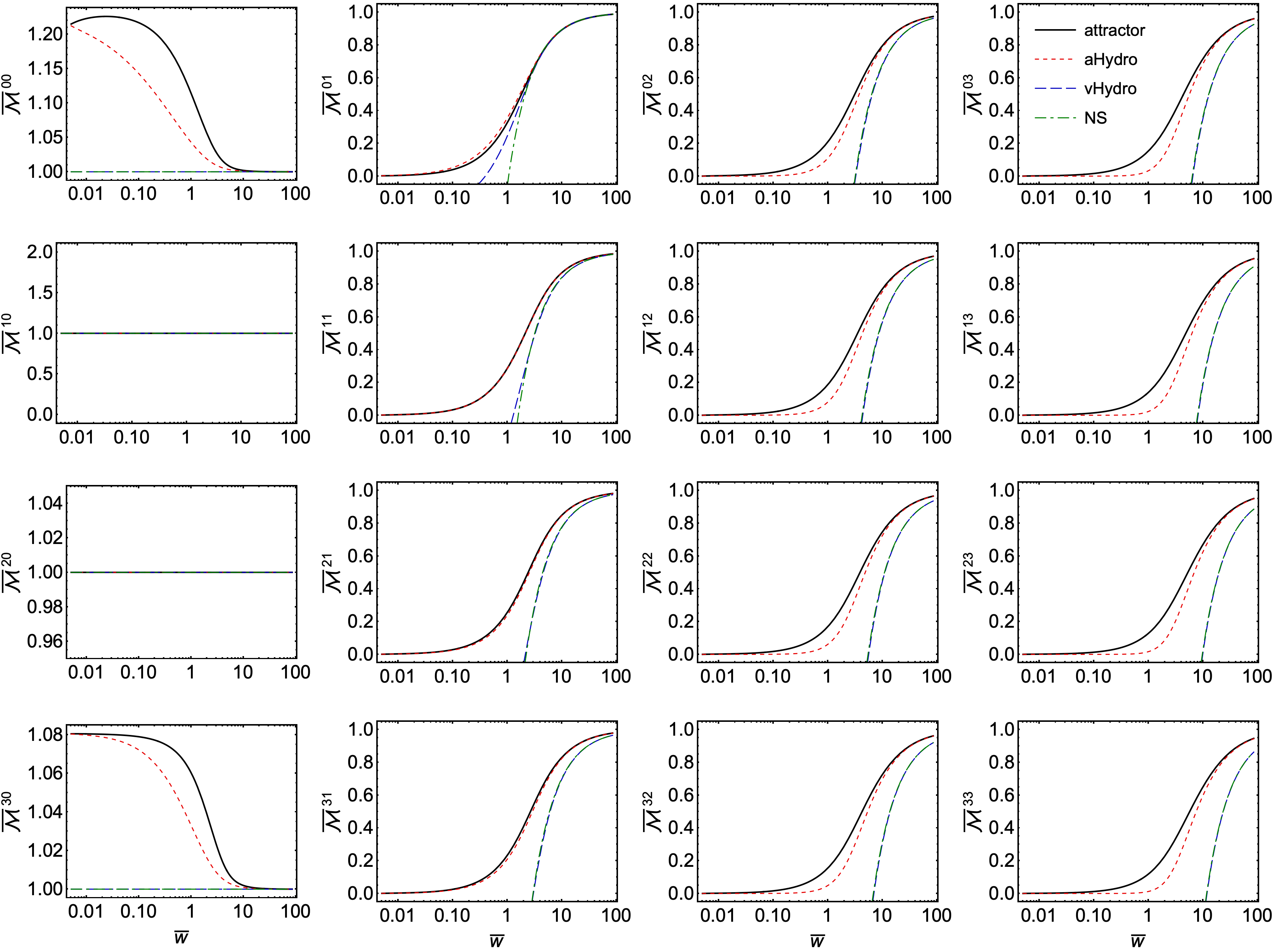}
}
\caption{Scaled moments $\barM^{nm}$ obtained from the exact attractor solution (solid black line) compared with the aHydro attractor (red dashed lines), DNMR attractor (blue long dashed lines), and the Navier-Stokes limit for each moment (green dot-dashed lines).  Horizontal axis is $\bar{w} \equiv \tau T/5 \bar\eta$.  Panels show a grid in $n$ and $m$.}
\label{fig:attractorGridComp}
\end{figure*}

For aHydro, one extracts the anisotropy parameter $\alpha(\tau)$ associated with the attractor solution.  Once this is determined one can use Eq. (4.4) of Ref.~\cite{Strickland:2018ayk}, modified to take into account finite fugacity, to obtain a compact expression for any moment
\be
\barM^{nm}_{\rm aHydro}(\tau) = (2m+1) (2 \alpha)^{n+2m-2}\frac{{\cal H}^{nm}(\alpha)}{[{\cal H}(\alpha)]^{n+2m-1}} \, .
\label{eq:ahydromoms}
\ee
\checked{u}
The aHydro dynamical equations taking into account number conservation can be found in Ref.~\cite{Almaalol:2018jmz}.  In Fig.~\ref{fig:attractorGridComp}, I compare the exact attractor (black solid lines) with the aHydro attractor (red dashed lines), DNMR attractor (blue long dashed lines), and the NS limit (green dot-dashed lines) for each moment.  In all cases shown, aHydro provides a better approximation to the exact moments than the vHydro or NS frameworks.  This is similar to what was found in the case where number conservation was not imposed \cite{Strickland:2018ayk}.  Finally, we note that, for aHydro, the moment with the best agreement is the $n=1$ and $m=1$ moment, for which the two results are virtually indistinguishable.  This can be contrasted with Ref.~\cite{Strickland:2018ayk} which found that it was the $n=0$ and $m=1$ moment which was best described when number conservation was not imposed.  It is not clear to us why this would be the case.

\section{Conclusions}
\label{sect:conclusions}

In this paper we extended previous studies of the conformal 0+1d kinetic non-equilibrium attractor in relaxation time approximation by imposing number conservation through the introduction of a dynamical fugacity (chemical potential).  We derived two coupled integral equations for the effective temperature and fugacity which were then solved numerically to obtain the exact solution.  We demonstrated that the resulting solutions exhibited convergence to a unique non-equilibrium attractor even though the system is out of chemical equilibrium generically ($\lim_{\tau \rightarrow \infty} \Gamma(\tau) \neq 1$).  We found that, compared to the case where number conservation was not imposed, the moments converge to their respective attractors more quickly.  Overall, however, we found that the behavior in the two cases is qualitatively very similar, providing further evidence that the non-equilibrium attractor is ubiquitous.  We also compared the resulting attractor moments with predictions of different hydrodynamic frameworks.  We found that anisotropic hydrodynamics provided the best approximation to the exact results for all moments.

Looking forward, herein we used a RTA collisional kernel and enforced number conservation by introducing a dynamical fugacity.  It would be interesting to look at leading-order scalar field theory, in which case one only has 2 $\leftrightarrow$ 2 collisions and hence a theory which automatically conserves number.  Such comparisons have been made in the context of aHydro in Ref.~\cite{Almaalol:2018jmz} where the authors studied both number-conserving RTA and scalar collisional kernels.  Therein, it was shown that one could numerically extract the aHydro attractor for both RTA and scalar kernels, with the two being qualitatively similar.  It would be very interesting to consider the 2 $\leftrightarrow$ 2 scalar kinetic theory using Monte-Carlo-based transport in order to compare with the exact results obtained herein using number-conserving RTA.  It would also be interesting to make comparisons with the attractor extracted from the effective kinetic theory framework of Kurkela et al, particularly in the case that baryon number conservation is at play~\cite{Kurkela:2015qoa,Keegan:2015avk,Kurkela:2018oqw,Kurkela:2018xxd}.

\acknowledgments

We thank D.~Almaalol for comments and suggestions.  M.S. and U.T. were supported by the U.S. Department of Energy, Office of Science, Office of Nuclear Physics under Award No. DE-SC0013470.

\bibliographystyle{JHEP}
\bibliography{ncrta}

\providecommand{\href}[2]{#2}\begingroup\raggedright\begin{thebibliography}{10}

\bibitem{Averbeck:2015jja}
R.~Averbeck, J.~W. Harris and B.~Schenke, \emph{{Heavy-Ion Physics at the
  LHC}},  in \emph{The Large Hadron Collider: Harvest of Run 1}
  (T.~Sch{\"o}rner-Sadenius, ed.), pp.~355--420.
\newblock 2015.
\newblock \href{https://doi.org/10.1007/978-3-319-15001-7_9}{DOI}.

\bibitem{Jeon:2016uym}
S.~Jeon and U.~Heinz, \emph{{Introduction to Hydrodynamics}},  in
  \emph{Quark-Gluon Plasma 5} (X.-N. Wang, ed.), pp.~131--187.
\newblock 2016.
\newblock \href{https://doi.org/10.1142/9789814663717_0003}{DOI}.

\bibitem{Romatschke:2017ejr}
P.~Romatschke and U.~Romatschke, \emph{{Relativistic Fluid Dynamics In and Out
  of Equilibrium -- Ten Years of Progress in Theory and Numerical Simulations
  of Nuclear Collisions}},  \href{https://arxiv.org/abs/1712.05815}{{\ttfamily
  1712.05815}}.

\bibitem{Florkowski:2017olj}
W.~Florkowski, M.~P. Heller and M.~Spalinski, \emph{{New theories of
  relativistic hydrodynamics in the LHC era}},
  \href{https://arxiv.org/abs/1707.02282}{{\ttfamily 1707.02282}}.

\bibitem{Alqahtani:2017mhy}
M.~Alqahtani, M.~Nopoush and M.~Strickland, \emph{{Relativistic anisotropic
  hydrodynamics}},
  \href{https://doi.org/10.1016/j.ppnp.2018.05.004}{\emph{Prog. Part. Nucl.
  Phys.} {\bfseries 101} (2018) 204}
  [\href{https://arxiv.org/abs/1712.03282}{{\ttfamily 1712.03282}}].

\bibitem{Chesler:2008hg}
P.~M. Chesler and L.~G. Yaffe, \emph{{Horizon formation and
  far-from-equilibrium isotropization in supersymmetric Yang-Mills plasma}},
  \href{https://doi.org/10.1103/PhysRevLett.102.211601}{\emph{Phys. Rev. Lett.}
  {\bfseries 102} (2009) 211601}
  [\href{https://arxiv.org/abs/0812.2053}{{\ttfamily 0812.2053}}].

\bibitem{Heller:2013oxa}
M.~P. Heller, D.~Mateos, W.~van~der Schee and M.~Triana, \emph{{Holographic
  isotropization linearized}},
  \href{https://doi.org/10.1007/JHEP09(2013)026}{\emph{JHEP} {\bfseries 09}
  (2013) 026} [\href{https://arxiv.org/abs/1304.5172}{{\ttfamily 1304.5172}}].

\bibitem{Keegan:2015avk}
L.~Keegan, A.~Kurkela, P.~Romatschke, W.~van~der Schee and Y.~Zhu, \emph{{Weak
  and strong coupling equilibration in nonabelian gauge theories}},
  \href{https://doi.org/10.1007/JHEP04(2016)031}{\emph{JHEP} {\bfseries 04}
  (2016) 031} [\href{https://arxiv.org/abs/1512.05347}{{\ttfamily
  1512.05347}}].

\bibitem{Romatschke:2017vte}
P.~Romatschke, \emph{{Relativistic Fluid Dynamics Far From Local Equilibrium}},
  \href{https://doi.org/10.1103/PhysRevLett.120.012301}{\emph{Phys. Rev. Lett.}
  {\bfseries 120} (2018) 012301}
  [\href{https://arxiv.org/abs/1704.08699}{{\ttfamily 1704.08699}}].

\bibitem{Strickland:2017kux}
M.~Strickland, J.~Noronha and G.~Denicol, \emph{{Anisotropic nonequilibrium
  hydrodynamic attractor}},
  \href{https://doi.org/10.1103/PhysRevD.97.036020}{\emph{Phys. Rev.}
  {\bfseries D97} (2018) 036020}
  [\href{https://arxiv.org/abs/1709.06644}{{\ttfamily 1709.06644}}].

\bibitem{Heller:2015dha}
M.~P. Heller and M.~Spalinski, \emph{{Hydrodynamics Beyond the Gradient
  Expansion: Resurgence and Resummation}},
  \href{https://doi.org/10.1103/PhysRevLett.115.072501}{\emph{Phys. Rev. Lett.}
  {\bfseries 115} (2015) 072501}
  [\href{https://arxiv.org/abs/1503.07514}{{\ttfamily 1503.07514}}].

\bibitem{Kurkela:2015qoa}
A.~Kurkela and Y.~Zhu, \emph{{Isotropization and hydrodynamization in weakly
  coupled heavy-ion collisions}},
  \href{https://doi.org/10.1103/PhysRevLett.115.182301}{\emph{Phys. Rev. Lett.}
  {\bfseries 115} (2015) 182301}
  [\href{https://arxiv.org/abs/1506.06647}{{\ttfamily 1506.06647}}].

\bibitem{JaiswalForth}
C.~Chattopadhyay, A.~Jaiswal, S.~Jaiswal and S.~Pal, \emph{Analytical solutions
  and attractors of higher-order viscous hydrodynamics for bjorken flow},
  {\emph{\rm forthcoming} (2019) }.

\bibitem{Behtash:2017wqg}
A.~Behtash, C.~N. Cruz-Camacho and M.~Martinez, \emph{{Far-from-equilibrium
  attractors and nonlinear dynamical systems approach to the Gubser flow}},
  \href{https://doi.org/10.1103/PhysRevD.97.044041}{\emph{Phys. Rev.}
  {\bfseries D97} (2018) 044041}
  [\href{https://arxiv.org/abs/1711.01745}{{\ttfamily 1711.01745}}].

\bibitem{Behtash:2018moe}
A.~Behtash, S.~Kamata, M.~Martinez and C.~N. Cruz-Camacho,
  \emph{{Non-perturbative rheological behavior of a far-from-equilibrium
  expanding plasma}},  \href{https://arxiv.org/abs/1805.07881}{{\ttfamily
  1805.07881}}.

\bibitem{Denicol:2018pak}
G.~S. Denicol and J.~Noronha, \emph{{Hydrodynamic attractor and the fate of
  perturbative expansions in Gubser flow}},
  \href{https://arxiv.org/abs/1804.04771}{{\ttfamily 1804.04771}}.

\bibitem{Strickland:2018ayk}
M.~Strickland, \emph{{The non-equilibrium attractor for kinetic theory in
  relaxation time approximation}},
  \href{https://doi.org/10.1007/JHEP12(2018)128}{\emph{JHEP} {\bfseries 12}
  (2018) 128} [\href{https://arxiv.org/abs/1809.01200}{{\ttfamily
  1809.01200}}].

\bibitem{Behtash:2019txb}
A.~Behtash, S.~Kamata, M.~Martinez and H.~Shi, \emph{{Dynamical systems and
  nonlinear transient rheology of the far-from-equilibrium Bjorken flow}},
  \href{https://arxiv.org/abs/1901.08632}{{\ttfamily 1901.08632}}.

\bibitem{Eckart:1940te}
C.~Eckart, \emph{{The Thermodynamics of irreversible processes. 3..
  Relativistic theory of the simple fluid}},
  \href{https://doi.org/10.1103/PhysRev.58.919}{\emph{Phys. Rev.} {\bfseries
  58} (1940) 919}.

\bibitem{ldlandau2013}
L.~D. Landau, \emph{Fluid Mechanics: Landau and Lifshitz: Course of Theoretical
  Physics, Volume 6}. Pergamon.

\bibitem{Weinberg:1971mx}
S.~Weinberg, \emph{{Entropy generation and the survival of protogalaxies in an
  expanding universe}}, \href{https://doi.org/10.1086/151073}{\emph{Astrophys.
  J.} {\bfseries 168} (1971) 175}.

\bibitem{Muller:1967zza}
I.~Muller, \emph{{Zum Paradoxon der Warmeleitungstheorie}},
  \href{https://doi.org/10.1007/BF01326412}{\emph{Z. Phys.} {\bfseries 198}
  (1967) 329}.

\bibitem{Israel:1976tn}
W.~Israel, \emph{{Nonstationary irreversible thermodynamics: A Causal
  relativistic theory}},
  \href{https://doi.org/10.1016/0003-4916(76)90064-6}{\emph{Annals Phys.}
  {\bfseries 100} (1976) 310}.

\bibitem{Israel:1979wp}
W.~Israel and J.~M. Stewart, \emph{{Transient relativistic thermodynamics and
  kinetic theory}},
  \href{https://doi.org/10.1016/0003-4916(79)90130-1}{\emph{Annals Phys.}
  {\bfseries 118} (1979) 341}.

\bibitem{Muronga:2001zk}
A.~Muronga, \emph{{Second order dissipative fluid dynamics for ultra-
  relativistic nuclear collisions}},
  \href{https://doi.org/10.1103/PhysRevLett.88.062302}{\emph{Phys. Rev. Lett.}
  {\bfseries 88} (2002) 062302}
  [\href{https://arxiv.org/abs/nucl-th/0104064}{{\ttfamily nucl-th/0104064}}].

\bibitem{Muronga:2003ta}
A.~Muronga, \emph{{Causal Theories of Dissipative Relativistic Fluid Dynamics
  for Nuclear Collisions}},
  \href{https://doi.org/10.1103/PhysRevC.69.034903}{\emph{Phys. Rev.}
  {\bfseries C69} (2004) 034903}
  [\href{https://arxiv.org/abs/nucl-th/0309055}{{\ttfamily nucl-th/0309055}}].

\bibitem{Muronga:2004sf}
A.~Muronga and D.~H. Rischke, \emph{{Evolution of hot, dissipative quark matter
  in relativistic nuclear collisions}},
  \href{https://arxiv.org/abs/nucl-th/0407114}{{\ttfamily nucl-th/0407114}}.

\bibitem{Heinz:2005bw}
U.~W. Heinz, H.~Song and A.~K. Chaudhuri, \emph{{Dissipative hydrodynamics for
  viscous relativistic fluids}},
  \href{https://doi.org/10.1103/PhysRevC.73.034904}{\emph{Phys.Rev.} {\bfseries
  C73} (2006) 034904} [\href{https://arxiv.org/abs/nucl-th/0510014}{{\ttfamily
  nucl-th/0510014}}].

\bibitem{Baier:2006um}
R.~Baier, P.~Romatschke and U.~A. Wiedemann, \emph{{Dissipative hydrodynamics
  and heavy ion collisions}},
  \href{https://doi.org/10.1103/PhysRevC.73.064903}{\emph{Phys.Rev.} {\bfseries
  C73} (2006) 064903} [\href{https://arxiv.org/abs/hep-ph/0602249}{{\ttfamily
  hep-ph/0602249}}].

\bibitem{Romatschke:2007mq}
P.~Romatschke and U.~Romatschke, \emph{{Viscosity Information from Relativistic
  Nuclear Collisions: How Perfect is the Fluid Observed at RHIC?}},
  \href{https://doi.org/10.1103/PhysRevLett.99.172301}{\emph{Phys. Rev. Lett.}
  {\bfseries 99} (2007) 172301}
  [\href{https://arxiv.org/abs/0706.1522}{{\ttfamily 0706.1522}}].

\bibitem{Baier:2007ix}
R.~Baier, P.~Romatschke, D.~T. Son, A.~O. Starinets and M.~A. Stephanov,
  \emph{{Relativistic viscous hydrodynamics, conformal invariance, and
  holography}},
  \href{https://doi.org/10.1088/1126-6708/2008/04/100}{\emph{JHEP} {\bfseries
  0804} (2008) 100} [\href{https://arxiv.org/abs/0712.2451}{{\ttfamily
  0712.2451}}].

\bibitem{Dusling:2007gi}
K.~Dusling and D.~Teaney, \emph{{Simulating elliptic flow with viscous
  hydrodynamics}},
  \href{https://doi.org/10.1103/PhysRevC.77.034905}{\emph{Phys. Rev.}
  {\bfseries C77} (2008) 034905}
  [\href{https://arxiv.org/abs/0710.5932}{{\ttfamily 0710.5932}}].

\bibitem{Luzum:2008cw}
M.~Luzum and P.~Romatschke, \emph{{Conformal Relativistic Viscous
  Hydrodynamics: Applications to RHIC results at sqrt(sNN) = 200 GeV}},
  \href{https://doi.org/10.1103/PhysRevC.78.034915}{\emph{Phys. Rev.}
  {\bfseries C78} (2008) 034915}
  [\href{https://arxiv.org/abs/0804.4015}{{\ttfamily 0804.4015}}].

\bibitem{Song:2008hj}
H.~Song and U.~W. Heinz, \emph{{Extracting the QGP viscosity from RHIC data - A
  Status report from viscous hydrodynamics}},
  \href{https://doi.org/10.1088/0954-3899/36/6/064033}{\emph{J.Phys.G}
  {\bfseries G36} (2009) 064033}
  [\href{https://arxiv.org/abs/0812.4274}{{\ttfamily 0812.4274}}].

\bibitem{Heinz:2009xj}
U.~W. Heinz, \emph{{Early collective expansion: Relativistic hydrodynamics and
  the transport properties of \protect{QCD} matter}}, {\emph{Relativistic Heavy
  Ion Physics, Landolt-Boernstein New Series, I/23, edited by R. Stock,
  Springer Verlag, New York, Chap. 5} (2010) }
  [\href{https://arxiv.org/abs/0901.4355}{{\ttfamily 0901.4355}}].

\bibitem{Schenke:2010rr}
B.~Schenke, S.~Jeon and C.~Gale, \emph{{Elliptic and triangular flow in
  event-by-event (3+1)D viscous hydrodynamics}},
  \href{https://doi.org/10.1103/PhysRevLett.106.042301}{\emph{Phys.Rev.Lett.}
  {\bfseries 106} (2011) 042301}
  [\href{https://arxiv.org/abs/1009.3244}{{\ttfamily 1009.3244}}].

\bibitem{Schenke:2011tv}
B.~Schenke, S.~Jeon and C.~Gale, \emph{{Anisotropic flow in sqrt(s)=2.76 TeV
  Pb+Pb collisions at the LHC}},
  \href{https://doi.org/10.1016/j.physletb.2011.06.065}{\emph{Phys.Lett.}
  {\bfseries B702} (2011) 59}
  [\href{https://arxiv.org/abs/1102.0575}{{\ttfamily 1102.0575}}].

\bibitem{Bozek:2011wa}
P.~Bozek, \emph{{Components of the elliptic flow in Pb-Pb collisions at
  s**(1/2) = 2.76-TeV}},
  \href{https://doi.org/10.1016/j.physletb.2011.04.020}{\emph{Phys.Lett.}
  {\bfseries B699} (2011) 283}
  [\href{https://arxiv.org/abs/1101.1791}{{\ttfamily 1101.1791}}].

\bibitem{Niemi:2011ix}
H.~Niemi, G.~S. Denicol, P.~Huovinen, E.~Moln\'{a}r and D.~H. Rischke,
  \emph{{Influence of the shear viscosity of the quark-gluon plasma on elliptic
  flow in ultrarelativistic heavy-ion collisions}},
  \href{https://doi.org/10.1103/PhysRevLett.106.212302}{\emph{Phys.Rev.Lett.}
  {\bfseries 106} (2011) 212302}
  [\href{https://arxiv.org/abs/1101.2442}{{\ttfamily 1101.2442}}].

\bibitem{Denicol:2011fa}
G.~S. Denicol, J.~Noronha, H.~Niemi and D.~H. Rischke, \emph{{Origin of the
  Relaxation Time in Dissipative Fluid Dynamics}},
  \href{https://doi.org/10.1103/PhysRevD.83.074019}{\emph{Phys. Rev.}
  {\bfseries D83} (2011) 074019}
  [\href{https://arxiv.org/abs/1102.4780}{{\ttfamily 1102.4780}}].

\bibitem{Niemi:2012ry}
H.~Niemi, G.~S. Denicol, P.~Huovinen, E.~Moln\'{a}r and D.~H. Rischke,
  \emph{Influence of a temperature-dependent shear viscosity on the azimuthal
  asymmetries of transverse momentum spectra in ultrarelativistic heavy-ion
  collisions}, \href{https://doi.org/10.1103/PhysRevC.86.014909}{\emph{Phys.
  Rev. C} {\bfseries 86} (2012) 014909}.

\bibitem{Bozek:2012qs}
P.~Bo\ifmmode~\dot{z}\else \.{z}\fi{}ek and I.~Wyskiel-Piekarska,
  \emph{Particle spectra in pb-pb collisions at
  $\sqrt{{\mathbf{s}}_{\mathbf{NN}}}=\mathbf{2}.\mathbf{76}$ tev},
  \href{https://doi.org/10.1103/PhysRevC.85.064915}{\emph{Phys. Rev. C}
  {\bfseries 85} (2012) 064915}.

\bibitem{Denicol:2012cn}
G.~S. Denicol, H.~Niemi, E.~Molnar and D.~H. Rischke, \emph{{Derivation of
  transient relativistic fluid dynamics from the Boltzmann equation}},
  \href{https://doi.org/10.1103/PhysRevD.85.114047,
  10.1103/PhysRevD.91.039902}{\emph{Phys. Rev.} {\bfseries D85} (2012) 114047}
  [\href{https://arxiv.org/abs/1202.4551}{{\ttfamily 1202.4551}}].

\bibitem{Denicol:2012es}
G.~Denicol, E.~Moln\'{a}r, H.~Niemi and D.~Rischke, \emph{{Derivation of fluid
  dynamics from kinetic theory with the 14--moment approximation}},
  \href{https://doi.org/10.1140/epja/i2012-12170-x}{\emph{Eur. Phys. J. A}
  {\bfseries 48} (2012) 170} [\href{https://arxiv.org/abs/1206.1554}{{\ttfamily
  1206.1554}}].

\bibitem{Denicol:2014vaa}
G.~S. Denicol, S.~Jeon and C.~Gale, \emph{{Transport Coefficients of Bulk
  Viscous Pressure in the 14-moment approximation}},
  \href{https://doi.org/10.1103/PhysRevC.90.024912}{\emph{Phys. Rev.}
  {\bfseries C90} (2014) 024912}
  [\href{https://arxiv.org/abs/1403.0962}{{\ttfamily 1403.0962}}].

\bibitem{Denicol:2014mca}
G.~S. Denicol, W.~Florkowski, R.~Ryblewski and M.~Strickland, \emph{{Shear-bulk
  coupling in nonconformal hydrodynamics}},
  \href{https://doi.org/10.1103/PhysRevC.90.044905}{\emph{Phys.Rev.} {\bfseries
  C90} (2014) 044905} [\href{https://arxiv.org/abs/1407.4767}{{\ttfamily
  1407.4767}}].

\bibitem{Jaiswal:2014isa}
A.~Jaiswal, R.~Ryblewski and M.~Strickland, \emph{{Transport coefficients for
  bulk viscous evolution in the relaxation time approximation}},
  \href{https://doi.org/10.1103/PhysRevC.90.044908}{\emph{Phys. Rev.}
  {\bfseries C90} (2014) 044908}
  [\href{https://arxiv.org/abs/1407.7231}{{\ttfamily 1407.7231}}].

\bibitem{Jaiswal:2013npa}
A.~Jaiswal, \emph{{Relativistic dissipative hydrodynamics from kinetic theory
  with relaxation time approximation}},
  \href{https://doi.org/10.1103/PhysRevC.87.051901}{\emph{Phys. Rev.}
  {\bfseries C87} (2013) 051901}
  [\href{https://arxiv.org/abs/1302.6311}{{\ttfamily 1302.6311}}].

\bibitem{Jaiswal:2013vta}
A.~Jaiswal, \emph{{Relativistic third-order dissipative fluid dynamics from
  kinetic theory}},
  \href{https://doi.org/10.1103/PhysRevC.88.021903}{\emph{Phys. Rev.}
  {\bfseries C88} (2013) 021903}
  [\href{https://arxiv.org/abs/1305.3480}{{\ttfamily 1305.3480}}].

\bibitem{Florkowski:2010cf}
W.~Florkowski and R.~Ryblewski, \emph{{Highly-anisotropic and
  strongly-dissipative hydrodynamics for early stages of relativistic heavy-ion
  collisions}},
  \href{https://doi.org/10.1103/PhysRevC.83.034907}{\emph{Phys.Rev.} {\bfseries
  C83} (2011) 034907} [\href{https://arxiv.org/abs/1007.0130}{{\ttfamily
  1007.0130}}].

\bibitem{Martinez:2010sc}
M.~Martinez and M.~Strickland, \emph{{Dissipative Dynamics of Highly
  Anisotropic Systems}},
  \href{https://doi.org/10.1016/j.nuclphysa.2010.08.011}{\emph{Nucl. Phys.}
  {\bfseries A848} (2010) 183}
  [\href{https://arxiv.org/abs/1007.0889}{{\ttfamily 1007.0889}}].

\bibitem{Ryblewski:2010ch}
R.~Ryblewski and W.~Florkowski, \emph{{Highly anisotropic hydrodynamics --
  discussion of the model assumptions and forms of the initial conditions}},
  \href{https://doi.org/10.5506/APhysPolB.42.115}{\emph{Acta Phys. Polon.}
  {\bfseries B42} (2011) 115}
  [\href{https://arxiv.org/abs/1011.6213}{{\ttfamily 1011.6213}}].

\bibitem{Florkowski:2011jg}
W.~Florkowski and R.~Ryblewski, \emph{{Projection method for boost-invariant
  and cylindrically symmetric dissipative hydrodynamics}},
  \href{https://doi.org/10.1103/PhysRevC.85.044902}{\emph{Phys.Rev.} {\bfseries
  C85} (2012) 044902} [\href{https://arxiv.org/abs/1111.5997}{{\ttfamily
  1111.5997}}].

\bibitem{Martinez:2012tu}
M.~Martinez, R.~Ryblewski and M.~Strickland, \emph{{Boost-Invariant
  (2+1)-dimensional Anisotropic Hydrodynamics}}, {\emph{Phys.Rev.} {\bfseries
  C85} (2012) 064913} [\href{https://arxiv.org/abs/1204.1473}{{\ttfamily
  1204.1473}}].

\bibitem{Ryblewski:2012rr}
R.~Ryblewski and W.~Florkowski, \emph{{Highly-anisotropic hydrodynamics in 3+1
  space-time dimensions}},
  \href{https://doi.org/10.1103/PhysRevC.85.064901}{\emph{Phys. Rev.}
  {\bfseries C85} (2012) 064901}
  [\href{https://arxiv.org/abs/1204.2624}{{\ttfamily 1204.2624}}].

\bibitem{Bazow:2013ifa}
D.~Bazow, U.~W. Heinz and M.~Strickland, \emph{{Second-order (2+1)-dimensional
  anisotropic hydrodynamics}},
  \href{https://doi.org/10.1103/PhysRevC.90.054910}{\emph{Phys.Rev.} {\bfseries
  C90} (2014) 054910} [\href{https://arxiv.org/abs/1311.6720}{{\ttfamily
  1311.6720}}].

\bibitem{Tinti:2013vba}
L.~Tinti and W.~Florkowski, \emph{{Projection method and new formulation of
  leading-order anisotropic hydrodynamics}},
  \href{https://doi.org/10.1103/PhysRevC.89.034907}{\emph{Phys.Rev.} {\bfseries
  C89} (2014) 034907} [\href{https://arxiv.org/abs/1312.6614}{{\ttfamily
  1312.6614}}].

\bibitem{Nopoush:2014pfa}
M.~Nopoush, R.~Ryblewski and M.~Strickland, \emph{{Bulk viscous evolution
  within anisotropic hydrodynamics}},
  \href{https://doi.org/10.1103/PhysRevC.90.014908}{\emph{Phys.Rev.} {\bfseries
  C90} (2014) 014908} [\href{https://arxiv.org/abs/1405.1355}{{\ttfamily
  1405.1355}}].

\bibitem{Florkowski:2014bba}
W.~Florkowski, R.~Ryblewski, M.~Strickland and L.~Tinti, \emph{{Leading-order
  anisotropic hydrodynamics for systems with massive particles}},
  \href{https://doi.org/10.1103/PhysRevC.89.054909}{\emph{Phys.Rev.} {\bfseries
  C89} (2014) 054909} [\href{https://arxiv.org/abs/1403.1223}{{\ttfamily
  1403.1223}}].

\bibitem{Tinti:2015xwa}
L.~Tinti, \emph{{Anisotropic matching principle for the hydrodynamic
  expansion}}, \href{https://doi.org/10.1103/PhysRevC.94.044902}{\emph{Phys.
  Rev.} {\bfseries C94} (2016) 044902}
  [\href{https://arxiv.org/abs/1506.07164}{{\ttfamily 1506.07164}}].

\bibitem{Bazow:2015cha}
D.~Bazow, U.~W. Heinz and M.~Martinez, \emph{{Nonconformal viscous anisotropic
  hydrodynamics}},
  \href{https://doi.org/http://dx.doi.org/10.1103/PhysRevC.91.064903}{\emph{Phys.Rev.}
  {\bfseries C91} (2015) 064903}
  [\href{https://arxiv.org/abs/1503.07443}{{\ttfamily 1503.07443}}].

\bibitem{Bazow:2015zca}
D.~Bazow, M.~Martinez and U.~W. Heinz, \emph{{Transient oscillations in a
  macroscopic effective theory of the Boltzmann equation}},
  \href{https://doi.org/10.1103/PhysRevD.93.034002}{\emph{Phys. Rev.}
  {\bfseries D93} (2016) 034002}
  [\href{https://arxiv.org/abs/1507.06595}{{\ttfamily 1507.06595}}].

\bibitem{Nopoush:2015yga}
M.~Nopoush, M.~Strickland, R.~Ryblewski, D.~Bazow, U.~Heinz and M.~Martinez,
  \emph{{Leading-order anisotropic hydrodynamics for central collisions}},
  \href{https://doi.org/10.1103/PhysRevC.92.044912}{\emph{Phys. Rev.}
  {\bfseries C92} (2015) 044912}
  [\href{https://arxiv.org/abs/1506.05278}{{\ttfamily 1506.05278}}].

\bibitem{Alqahtani:2015qja}
M.~Alqahtani, M.~Nopoush and M.~Strickland, \emph{{Quasiparticle equation of
  state for anisotropic hydrodynamics}},
  \href{https://doi.org/10.1103/PhysRevC.92.054910}{\emph{Phys. Rev.}
  {\bfseries C92} (2015) 054910}
  [\href{https://arxiv.org/abs/1509.02913}{{\ttfamily 1509.02913}}].

\bibitem{Molnar:2016vvu}
E.~Molnar, H.~Niemi and D.~H. Rischke, \emph{{Derivation of anisotropic
  dissipative fluid dynamics from the Boltzmann equation}},
  \href{https://doi.org/10.1103/PhysRevD.93.114025}{\emph{Phys. Rev.}
  {\bfseries D93} (2016) 114025}
  [\href{https://arxiv.org/abs/1602.00573}{{\ttfamily 1602.00573}}].

\bibitem{Molnar:2016gwq}
E.~Molnar, H.~Niemi and D.~H. Rischke, \emph{{Closing the equations of motion
  of anisotropic fluid dynamics by a judicious choice of a moment of the
  Boltzmann equation}},
  \href{https://doi.org/10.1103/PhysRevD.94.125003}{\emph{Phys. Rev.}
  {\bfseries D94} (2016) 125003}
  [\href{https://arxiv.org/abs/1606.09019}{{\ttfamily 1606.09019}}].

\bibitem{Bluhm:2015raa}
M.~Bluhm and T.~Schaefer, \emph{{Dissipative fluid dynamics for the dilute
  Fermi gas at unitarity: Anisotropic fluid dynamics}},
  \href{https://doi.org/10.1103/PhysRevA.92.043602}{\emph{Phys. Rev.}
  {\bfseries A92} (2015) 043602}
  [\href{https://arxiv.org/abs/1505.00846}{{\ttfamily 1505.00846}}].

\bibitem{Bluhm:2015bzi}
M.~Bluhm and T.~Schaefer, \emph{{Model-independent determination of the shear
  viscosity of a trapped unitary Fermi gas: Application to high temperature
  data}}, \href{https://doi.org/10.1103/PhysRevLett.116.115301}{\emph{Phys.
  Rev. Lett.} {\bfseries 116} (2016) 115301}
  [\href{https://arxiv.org/abs/1512.00862}{{\ttfamily 1512.00862}}].

\bibitem{Alqahtani:2017jwl}
M.~Alqahtani, M.~Nopoush, R.~Ryblewski and M.~Strickland, \emph{{(3+1)D
  Quasiparticle Anisotropic Hydrodynamics for Ultrarelativistic Heavy-Ion
  Collisions}},
  \href{https://doi.org/10.1103/PhysRevLett.119.042301}{\emph{Phys. Rev. Lett.}
  {\bfseries 119} (2017) 042301}
  [\href{https://arxiv.org/abs/1703.05808}{{\ttfamily 1703.05808}}].

\bibitem{Alqahtani:2017tnq}
M.~Alqahtani, M.~Nopoush, R.~Ryblewski and M.~Strickland, \emph{{Anisotropic
  hydrodynamic modeling of 2.76 TeV Pb-Pb collisions}},
  \href{https://doi.org/10.1103/PhysRevC.96.044910}{\emph{Phys. Rev.}
  {\bfseries C96} (2017) 044910}
  [\href{https://arxiv.org/abs/1705.10191}{{\ttfamily 1705.10191}}].

\bibitem{Almaalol:2018ynx}
D.~Almaalol and M.~Strickland, \emph{{Anisotropic hydrodynamics with a scalar
  collisional kernel}},
  \href{https://doi.org/10.1103/PhysRevC.97.044911}{\emph{Phys. Rev.}
  {\bfseries C97} (2018) 044911}
  [\href{https://arxiv.org/abs/1801.10173}{{\ttfamily 1801.10173}}].

\bibitem{Almaalol:2018gjh}
D.~Almaalol, M.~Alqahtani and M.~Strickland, \emph{{Anisotropic hydrodynamic
  modeling of 200 GeV Au-Au collisions}},
  \href{https://arxiv.org/abs/1807.04337}{{\ttfamily 1807.04337}}.

\bibitem{Florkowski:2012as}
W.~Florkowski, R.~Maj, R.~Ryblewski and M.~Strickland, \emph{{Hydrodynamics of
  anisotropic quark and gluon fluids}},
  \href{https://doi.org/10.1103/PhysRevC.87.034914}{\emph{Phys.Rev.} {\bfseries
  C87} (2013) 034914} [\href{https://arxiv.org/abs/1209.3671}{{\ttfamily
  1209.3671}}].

\bibitem{Florkowski:2015cba}
W.~Florkowski, E.~Maksymiuk, R.~Ryblewski and L.~Tinti, \emph{{Anisotropic
  hydrodynamics for mixture of quark and gluon fluids}},
  \href{https://arxiv.org/abs/1508.04534}{{\ttfamily 1508.04534}}.

\bibitem{Florkowski:2017ovw}
W.~Florkowski, E.~Maksymiuk and R.~Ryblewski, \emph{{Anisotropic-hydrodynamics
  approach to a quark-gluon fluid mixture}},
  \href{https://doi.org/10.1103/PhysRevC.97.014904}{\emph{Phys. Rev.}
  {\bfseries C97} (2018) 014904}
  [\href{https://arxiv.org/abs/1711.03872}{{\ttfamily 1711.03872}}].

\bibitem{Almaalol:2018jmz}
D.~Almaalol, M.~Alqahtani and M.~Strickland, \emph{{Anisotropic hydrodynamics
  with number-conserving kernels}},
  \href{https://doi.org/10.1103/PhysRevC.99.014903}{\emph{Phys. Rev.}
  {\bfseries C99} (2019) 014903}
  [\href{https://arxiv.org/abs/1808.07038}{{\ttfamily 1808.07038}}].

\bibitem{Bjorken:1982qr}
J.~D. Bjorken, \emph{{Highly Relativistic Nucleus-Nucleus Collisions: The
  Central Rapidity Region}},
  \href{https://doi.org/10.1103/PhysRevD.27.140}{\emph{Phys. Rev.} {\bfseries
  D27} (1983) 140}.

\bibitem{Bialas:1984wv}
A.~Bia\l{}as and W.~Czy\ifmmode~\dot{z}\else \.{z}\fi{}, \emph{Boost-invariant
  boltzmann-vlasov equations for relativistic quark-antiquark plasma},
  \href{https://doi.org/10.1103/PhysRevD.30.2371}{\emph{Phys. Rev. D}
  {\bfseries 30} (1984) 2371}.

\bibitem{Bialas:1987en}
A.~Bia\l{}as and W.~Czy\ifmmode~\dot{z}\else \.{z}\fi{}, \emph{Oscillations of
  quark-gluon plasma generated in strong color fields},
  \href{https://doi.org/10.1016/0550-3213(88)90035-1}{\emph{Nuclear Physics B}
  {\bfseries 296} (1988) 611 }.

\bibitem{Florkowski:2013lza}
W.~Florkowski, R.~Ryblewski and M.~Strickland, \emph{{Anisotropic Hydrodynamics
  for Rapidly Expanding Systems}},
  \href{https://doi.org/10.1016/j.nuclphysa.2013.08.004}{\emph{Nucl. Phys.}
  {\bfseries A916} (2013) 249}
  [\href{https://arxiv.org/abs/1304.0665}{{\ttfamily 1304.0665}}].

\bibitem{Florkowski:2013lya}
W.~Florkowski, R.~Ryblewski and M.~Strickland, \emph{{Testing viscous and
  anisotropic hydrodynamics in an exactly solvable case}},
  \href{https://doi.org/10.1103/PhysRevC.88.024903}{\emph{Phys. Rev.}
  {\bfseries C88} (2013) 024903}
  [\href{https://arxiv.org/abs/1305.7234}{{\ttfamily 1305.7234}}].

\bibitem{Romatschke:2003ms}
P.~Romatschke and M.~Strickland, \emph{{Collective modes of an anisotropic
  quark gluon plasma}},
  \href{https://doi.org/10.1103/PhysRevD.68.036004}{\emph{Phys. Rev.}
  {\bfseries D68} (2003) 036004}
  [\href{https://arxiv.org/abs/hep-ph/0304092}{{\ttfamily hep-ph/0304092}}].

\bibitem{MikeCodeDB}
M.~Strickland. \url{http://personal.kent.edu/~mstrick6/code/}, 2017.

\bibitem{Mauricio:2007vz}
M.~Martinez and M.~Strickland, \emph{{Measuring QGP thermalization time with
  dileptons}},
  \href{https://doi.org/10.1103/PhysRevLett.100.102301}{\emph{Phys. Rev. Lett.}
  {\bfseries 100} (2008) 102301}
  [\href{https://arxiv.org/abs/0709.3576}{{\ttfamily 0709.3576}}].

\bibitem{Martinez:2008di}
M.~Martinez and M.~Strickland, \emph{{Pre-equilibrium dilepton production from
  an anisotropic quark-gluon plasma}},
  \href{https://doi.org/10.1103/PhysRevC.78.034917}{\emph{Phys.Rev.} {\bfseries
  C78} (2008) 034917} [\href{https://arxiv.org/abs/0805.4552}{{\ttfamily
  0805.4552}}].

\bibitem{Martinez:2009mf}
M.~Martinez and M.~Strickland, \emph{{Constraining relativistic viscous
  hydrodynamical evolution}},
  \href{https://doi.org/10.1103/PhysRevC.79.044903}{\emph{Phys. Rev.}
  {\bfseries C79} (2009) 044903}
  [\href{https://arxiv.org/abs/0902.3834}{{\ttfamily 0902.3834}}].

\bibitem{Denicol:2010xn}
G.~Denicol, T.~Koide and D.~Rischke, \emph{{Dissipative relativistic fluid
  dynamics: a new way to derive the equations of motion from kinetic theory}},
  \href{https://doi.org/10.1103/PhysRevLett.105.162501}{\emph{Phys.Rev.Lett.}
  {\bfseries 105} (2010) 162501}
  [\href{https://arxiv.org/abs/1004.5013}{{\ttfamily 1004.5013}}].

\bibitem{Kurkela:2018oqw}
A.~Kurkela and A.~Mazeliauskas, \emph{{Chemical equilibration in weakly coupled
  QCD}},  \href{https://arxiv.org/abs/1811.03068}{{\ttfamily 1811.03068}}.

\bibitem{Kurkela:2018xxd}
A.~Kurkela and A.~Mazeliauskas, \emph{{Chemical equilibration in hadronic
  collisions}},  \href{https://arxiv.org/abs/1811.03040}{{\ttfamily
  1811.03040}}.

\end{thebibliography}\endgroup

\end{document}